# Super-resolution technique enabled by higher-order Bessel beams in nanoscale optical structures

**APOORVA D, M UMMAL MOMEEN** *

*Magnetic Instrumentation and Applied Optics Laboratory, Department of Physics, School of Advanced Sciences, Vellore Institute of Technology, Vellore, Tamil Nadu, 632014, India.*
*[*ummalmomeen@gmail.com](mailto:ummalmomeen@gmail.com)*

In conventional optical microscopy, accessing subwavelength information is classically constrained by the Abbe diffraction limit which remains a significant challenge in nanophotonics. To overcome this limitation, we introduce a scheme that generates a tightly confined near-field focal spot under non-diffracting Bessel beam illumination, employing a series of high refractive-index nanoscale structures. Furthermore, the utilization of higher-order Bessel beams enables additional resolution enhancement beyond the diffraction limit. Our results demonstrate that the pyramidal nanostructure can generate a highly localized near-field confinement with a characteristic width of 10 nm, enabling subwavelength optical intensity localization. To elucidate the factors governing the localization behaviour, we systematically investigate the influence of different geometrical parameters related to the incident Bessel beam on the near field localization and shows that these factors critically determine the stability and spatial profile of the focal spot. These findings highlight the potential of the proposed approach for various applications in super-resolution techniques and quantum technology.

## 1. INTRODUCTION

Light microscopy is an indispensable tool in biological and nanophotonics research owing to its ability to provide non-invasive spatial dimensional imaging of intracellular structures. However, the spatial resolution of classical optical microscopes is fundamentally restricted by Abbe's diffraction limit [1]. Various super-resolution techniques have emerged as a potential alternative for enabling visualization beyond the classical diffraction limit in order to analyse nano range structures and living organisms more precisely [2]. Nanoscale photonic structures are good candidate for enhancing the coupling of light matter interaction to overcome diffraction-limited resolution which is pivotal for high-precision applications in biomedicine, materials science, microfluidics, and photonics [3].

In practice, the performance of traditional optical microscopy is hindered in capturing subwavelength information, since evanescent waves carrying high-spatial-frequency components which decay exponentially away from the sample surface, thereby preventing nanoscale features being resolved at the image plane [4]. Immersion microscopy breaks the long-standing barrier of capturing evanescent wave owing to the reduction of the effective wavelength to $\lambda/n$ in a medium of refractive index $n$. However, the achievable resolution enhancement is still constrained by several factors such as the limited range of refractive indices available in transparent immersion media, rapid decay and coupling efficiency [5].

Over the years, structured illumination microscopy (SIM) [6], single-molecule localization techniques such as photoactivated localization microscopy (PALM) and stochastic optical reconstruction microscopy (STORM) [7,8], plasmonic superlenses [9], and metamaterial-based systems [10] have significantly advanced sub-diffraction imaging. However, SIM suffers from limited imaging depth in highly scattering samples due to degradation of the structured illumination pattern, while its reconstruction process remains sensitive to noise, reducing image fidelity under low signal-to-noise conditions. PALM/STORM techniques require long acquisition times, high-intensity excitation, and computationally intensive image reconstruction methods. Plasmonic superlenses are constrained by severe metallic losses, limited penetration depth, and near-field operation. Similarly, metamaterial-based systems face challenges including narrow operational bandwidth, intrinsic dissipative losses, and complex nanoscale fabrication requirements. There are several potential alternative approaches for high resolution methods such as near field localization to detect the evanescent wave via Near-field scanning optical microscopy (NSOM) techniques [11], nonlinear far-field techniques, far field superlens and metamaterial-based lens structures to transfer evanescent wave to propagating wave [12]. The resolution efficiency at the far field techniques suffers from fabrication complications such as these designs involve multilayer stack with suitable cavity incorporation. At the same time, near field approaches include metallic structures to utilize localized plasmon resonance to detect evanescent wave which can damage biological cells due to heating [13] and also limited by slow detection mechanism [14]. More recently, subwavelength confinement has also been achieved through spatial frequency engineering using structured beams [15,16], super-oscillatory lenses, meta surfaces, and nanoscale phase-modulating elements. These approaches demonstrate that by exploiting near-field effects, nonlinear processes, or tailored wavefront modulation, optical fields can be confined well below the conventional diffraction limit.

In this work, we introduce an alternative approach to enhance the near field confinement in the subwavelength region by coupling a Bessel beam within a high index nanostructures. Bessel beams have attracted significant attention for enabling super-resolution imaging beyond the diffraction limit due to their non-diffracting propagation characteristics. Previous studies have explored the use of axicon lenses to control the spatial distribution of light in optical systems [17]. The inclusion of nanostructures modifies the spatial distribution of the incident optical field, enabling strong near-field localization within a subwavelength region [18].

Here we report a novel super-resolution technique based on three-dimensional optical nanostructures with diverse geometries, including axicon, cylindrical, conical, pyramidal structures, as well as solid immersion lens (SIL) based designs, under structured beam illumination with higher-order Bessel beams up to an order of three to suppress the classical diffraction barrier. The effect of higher order Bessel beam for the near field coupling on different nanostructures is computed and assessed the performance of the solid immersion lens (SIL) in order to improve near field confinement at the plane of interest (POI) position. Geometrical optimization is performed to achieve a single dominant intensity lobe by suppressing the dominating sidelobes. These near field coupling with photonic structures has potential applications towards optical trapping, optical manipulation and quantum cryptography.

## 2. NUMERICAL MODELLING METHODS

The fundamental Gaussian beam with beam waist $w_0$ exhibits diffraction-limited broadening during propagation. This limitation can be mitigated by employing diffraction-free Bessel beams, which are exact solutions of the Helmholtz equation and exhibit diffraction-free propagation over a finite distance, enabling a narrow central intensity spot [19]. We have considered a plane Gaussian wave function to couple with a high index nanoaxicon in order to generate a tiny near field focal spot to mimic the experimental feasibility of the simulation. It is known that by sending Gaussian beams through axicon can produce Bessel beam effectively. We also created Bessel beams through mathematical expression to compare the purity of the beam with the axicon structure. Schematic representation of these protocols are shown in Figure 1.

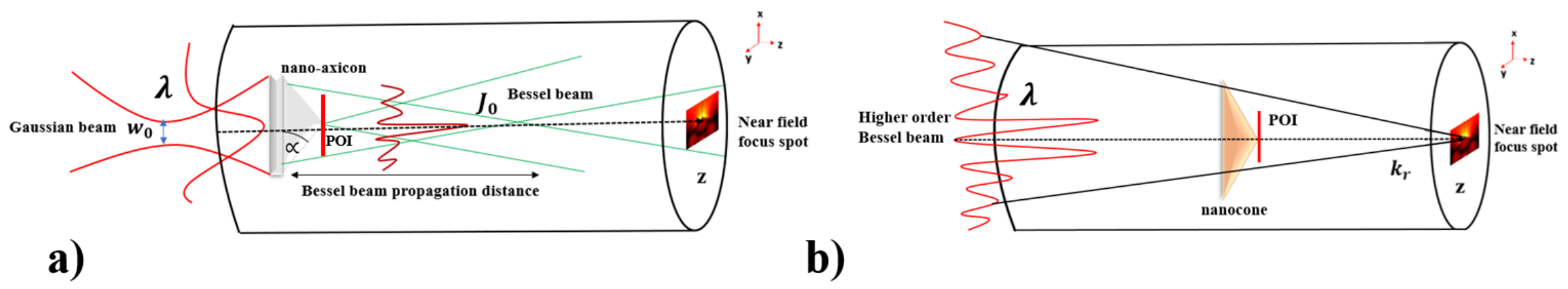

**Fig.1** Schematic representation for the generation of Bessel beam is shown here (a) Incident Gaussian beam in the presence of nano axicon to generate Bessel beam (b) Direct utilization of higher order Bessel beam field distribution equation under nano cone and other nanostructures.

The overall numerical modelling is executed using a finite element method (FEM) to solve Maxwell's equations in order to study the near field distribution via COMSOL Multiphysics software package. We have adopted a highly refined mesh with a maximum mesh size of $\lambda/5.05$ throughout the structure and minimum mesh size of $\lambda/125$ at the interfaces and edges to discretize the computational domain, ensuring accurate convergence of the electromagnetic field solution for the proposed nanoscale device. A cylindrical simulation area is considered to launch an incident beam with an aperture of 0.31. First, a classical approach is adopted to generate a Bessel beam using a high index (n=2.425) nanoaxicon, under a fundamental Gaussian beam illumination. We initialize our computation with a paraxial approximation of electric field distribution of the incident Gaussian beam which is expressed as $\mathrm{E}(\mathrm{x},\mathrm{y},\mathrm{z}) = \mathrm{A}_0 \frac{w_0}{\mathrm{w}(\mathrm{z})} e^{\left[-\frac{(x^2+y^2)}{w^2(z)}\right]} e^{-i(kz-\omega t)} e^{\left[-\frac{ik(x^2+y^2)}{2R(z)}\right]} e^{i\eta(z)}$, where $w_0$ is the beam waist and $\eta(z)$ represents the Gouy phase. Upon transmission through the nano-axicon, the Gaussian beam is transformed into a zeroth-order Bessel beam due to conical wavefront refraction and yields a non-diffracting central lobe and enhanced near-field confinement.

In another approach, Bessel beam is directly launched, described by a generalized solution of the Helmholtz equation incorporating Bessel functions $\mathrm{E}(\mathrm{x},\mathrm{y},\mathrm{z}) = \mathrm{A}_0 \mathrm{J}_m(\mathrm{k}_r r)\, e^{i k_z z}\, e^{i m \phi}$, where $\mathrm{J}_m$ is the Bessel function for order m. The notations $\mathrm{J}_0$, $\mathrm{J}_1$, $\mathrm{J}_2$, and $\mathrm{J}_3$ represent the zeroth, first, second, and third order Bessel beams, respectively, with a cone angle fixed at $\theta = 10^0$. $k_r = 2.05 \times 10^6\ m^{-1}$ and $k_z = 1.16 \times 10^7\ m^{-1}$ are the radial and longitudinal wave vector components respectively.

A numerical investigation was performed using different optical nanostructures under structured Bessel-beam illumination. Scattering boundary conditions were applied at the simulation boundaries to minimize unwanted reflections. A systematic three-dimensional geometrical parameter sweep was conducted, initially optimizing the structure under fundamental-mode illumination and subsequently investigating higher-order Bessel beams. The geometrical parameters were varied iteratively to identify configurations yielding highly localized near-field spots with full width at half maximum (FWHM) below 100 nm. The influence of individual geometrical parameters was monitored to identify the configurations providing optimal near-field confinement. The plane correspond to the top surface of nanostructure is considered as the plane of interest (POI) and serves as the reference for characterizing near-field confinement after the incident beam propagates through the nanostructure. Numerical simulations were performed for different Bessel-beam orders with optimized geometrical parameters of each nanostructure to achieve highly localized near-field spots while minimizing sidelobes. However, only the best near-field spot generation for each proposed nanostructure are included for the discussion. For $m = 0$, a strong on-axis intensity maximum is obtained, whereas $m \geq 1$ produces a central phase singularity and annular intensity distribution characteristics of optical vortex beams carrying orbital angular momentum. Increasing the beam order modifies the transverse spatial frequency content and can improve localization, but generally increases side lobe energy, leading to a trade-off between resolution and image contrast [20,21].

For consistent comparison, we computed the normalized near electric field distribution and the spatial confinement and it was quantified using the full width at half maximum (FWHM) of the corresponding intensity profile in the YZ plane. Specifically, the total electric-field intensity, $|E^2|$ was evaluated along the y-axis at x=0 on the selected POI. The FWHM was calculated as the peak intensity and its half-maximum value were first identified from the intensity profile along the (y)-axis at (x=0). The FWHM was then calculated along the (x)-axis. The same procedure was applied consistently to all investigated Bessel-beam orders and nanostructure geometries. A smaller FWHM indicates stronger lateral near-field confinement which can be potentially utilized for near field scanning over a substrate, yielding a better resolved information. The resultant localized fields may have potential applications in scanning-based imaging and nanoscale defect detection.

# 3. RESULTS AND DISCUSSION

## 3.1 Transformation of Gaussian beam into zeroth order Bessel beam by a nano-axicon

Figure 2 (a) illustrate the transformation of a fundamental Gaussian beam into a zeroth-order Bessel beam by utilizing nano-axicon structure. A monochromatic Gaussian beam with $\lambda = 532$ nm and beam waist $w_0 = 600$ nm propagating along the axial direction undergoes conical phase modulation upon transmission through a nano-axicon, acquiring a radial phase term $\exp(ik_r r)$. The subsequent axial interference of these conical wavefronts generates a Bessel-like field described by $J_0(k_r r)$, enabling non-diffracting propagation and enhanced spatial confinement. The optimized diamond based (n=2.425) nano-axicon cylindrical pedestal with the cone radius (R) =0.53 μm, base height (h) = 0.1 μm, cone height ($h_1$) = 0.215 μm and a conical angle ($\propto$) of 22° is being considered in this case.

Our computational study reveals that the constructive axial interference in the near-field which produces an ultra-small focal spot at the tip of the nano-axicon. We have considered our plane of interest (POI) at the tip of nano-axicon and obtained FWHM of 50 nm ($\approx \lambda/10.5$) (figure 2(b)) which indicates resolution of sub-diffraction-limited confinement resulting from near-field coupling and enhanced light concentration at the nano-axicon tip. In contrast, we have shown the near field coupling performance with a low-index silica axicon ($n = 1.45$) which yields significantly weaker confinement of FWHM =231 nm. It highlights the importance of refractive index contrast, hence materials like diamond and $TiO_2$ are essential for sustaining tight confinement, whereas low-index materials result in increased divergence and degraded focusing performance.

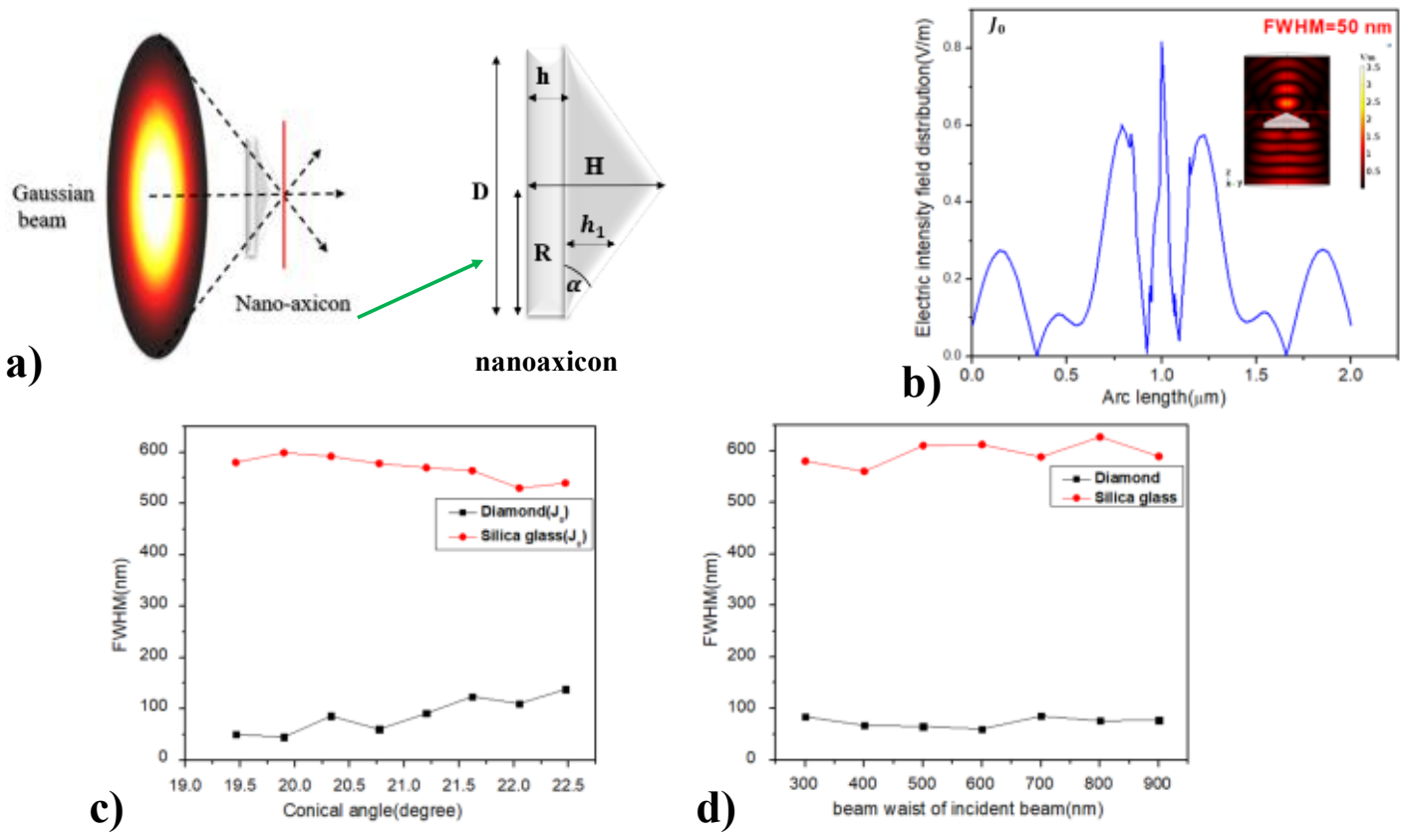

**Fig.2** (a) Schematic of Gaussian to zeroth order ($J_0$) Bessel beam transformation and geometrical configuration of the nano-axicon. (b) FEM simulated focal spot with corresponding spatial intensity distribution (inset) and extracted FWHM is shown here. (c) and (d) Dependence of FWHM on substrate refractive index, incident beam waist and nano-axicon geometrical parameters are shown.

To evaluate the robustness of the nanoaxicon design, a parametric sweep was performed by varying the geometrical and optical parameters, namely the conical angle, and incident beam waist. Figure 2(c) shows a stronger dependence on the conical angle ($\propto$), where variations between 20° and 22° lead to noticeable changes in focal spot size, highlighting the importance of axicon geometry in determining the beam confinement. The influence of the incident beam waist is presented in Figure 2(d) where the FWHM approaches a nearly constant value for beam waists within ~600 nm variation window, suggesting that sufficient illumination aperture is available for efficient Gaussian-to-Bessel beam conversion. Overall, the parametric study confirms stable sub-diffraction focusing over a broad range of design parameters, demonstrating the robustness of the proposed nanoaxicon configuration.

## 3.2 Subwavelength Optical Confinement with higher-order Bessel beam in nano-axicon structures

Figure 3 (a) shows the direct generation of zeroth and first order Bessel beams which is incident on a diamond-based nano-axicon in our simulation environment, where conical phase modulation reshapes the incident field into a Bessel-type distribution through enhancement of the transverse wave vector component $k_r$. Evanescent contributions and near-field coupling play a crucial role in determining field confinement at nanoscale geometrical configurations. We have coupled the first few modes (i.e. $J_0$, $J_1$, $J_2$, and $J_3$) of Bessel functions with our optimized nano-axicon structure, and achieved a highly resolved near field focal spot at $J_0$, and $J_1$ which are shown in Figure 3 (b). For $J_0$ mode, the optimized structure consists of a radius R = 0.58 µm, cone height ($h_1$) = 0.22 µm. For $J_1$, a slight geometrical modification is made in cone height ($h_1$= 0.215 µm) of the axicon to reduce the sidelobe intensities for higher orders. Strong scattering at the tip and edges redistributes energy into forward and lateral channels, while partial reflections induce axial interference, collectively defining the focal path for $J_1$ order. Our optimized configurations achieve FWHM values of 70 nm for $J_0$ and 165 nm for $J_1$ at the POI which is marked as red line at the tip of nano-axicon. Although higher-order modes reduce sidelobe intensity, they exhibit broader central features due to energy redistribution and the dependence of resolution on $k_r$. Figure 3 (c) indicates the FWHM variation when we vary our incident beam waist and the conical angle of nano-axicon for a Bessel mode $J_0$. Here we have considered different incident beam waists (i.e. $w_0$ = 300–900 nm) of higher order Bessel beam in order to check the stability of the near field coupling performance. We achieved a higher stability within a wider window of incident beam waist variation which validates that our optimized model is efficient. Also, we varied the conical angle (α) from 20°–22° with different beam waists for $J_0$, we have performed the same analysis for the next optimized mode (i.e. $J_1$) and our model is capable of generating a transcendent near field localization stability when we change the conical angle (α) within a range of 20°-22° and a beam waist variation of $w_0$ = 600–800 nm, which is depicted in Figure 3 (d). The low-index silica axicon (n = 1.45) highlighting the importance of refractive-index contrast for efficient light concentration and enhanced beam confinement.

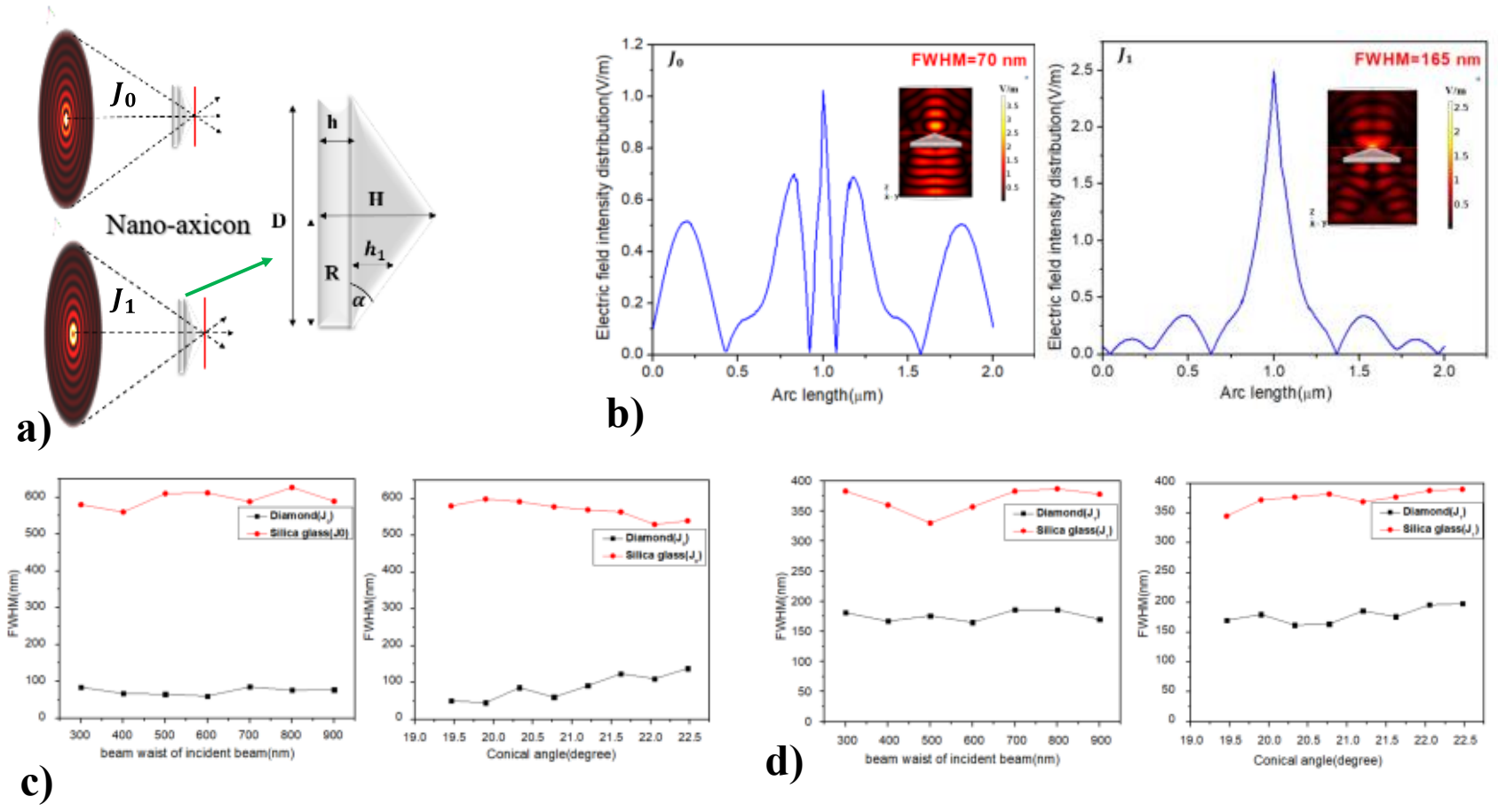

**Fig.3** (a) Schematic representation of nano-axicon focusing under direct Bessel beam illumination for zeroth ($J_0$) and first order ($J_1$) modes and representation of optimized structure defined by radius (R), height (H), and conical angle (α). (b) Spatial intensity distributions for $J_0$ and $J_1$ modes with corresponding FWHM comparison. (c), (d) Beam waist variation and conical angle dependence are shown for $J_0$ and $J_1$ beam excitation.

### 3.3 Tailoring near-field localization in nanocylinder structure via higher-order Bessel beam illumination

Figure 4(a) depicts the Bessel beams of order $J_0$ and $J_2$ which are illuminated through a high-index (diamond, n = 2.42 at 532 nm) nanocylinder. In this case, we achieved a better FWHM values only for $J_0$ and $J_2$ modes with our optimized structure (H = 0.615 µm, R = 0.150 µm) due to on-axis near field confinement. Under $J_0$ illumination, constructive on-axis interference produces a tightly confined central focal spot. Transitioning to higher-order Bessel modes alter the field symmetry and shifts the optical energy toward the nanocylinder boundaries. For the $J_2$ mode, this results in a pair of intense edge-localized focal spots with a reduced FWHM of 60 nm as shown in Figure 4 (b), suggesting enhanced subwavelength confinement through boundary-mediated scattering and cavity-like resonant field localization.

Structural tolerance is performed by varying the radius of the nano-cylinder from 140 nm to165 nm as shown in figures 4 (c) and 4 (d) and we achieved our desired FWHM i.e.148 nm and 60 nm for $J_0$ and $J_2$ orders respectively at the radius of 150 nm. Incident beam waist variations are also shown in figures 4(c) and 4(d). Simultaneously we have evaluated the FWHM values for the same nanostructures with commonly available lens material such as silica. This silica nano-cylinder weakens the near field confinement, consistent with effective wavelength scaling. Thus high-index nanocylinder acts as a subwavelength optical antenna, concentrating Bessel beam energy into localized near-field focal spots.

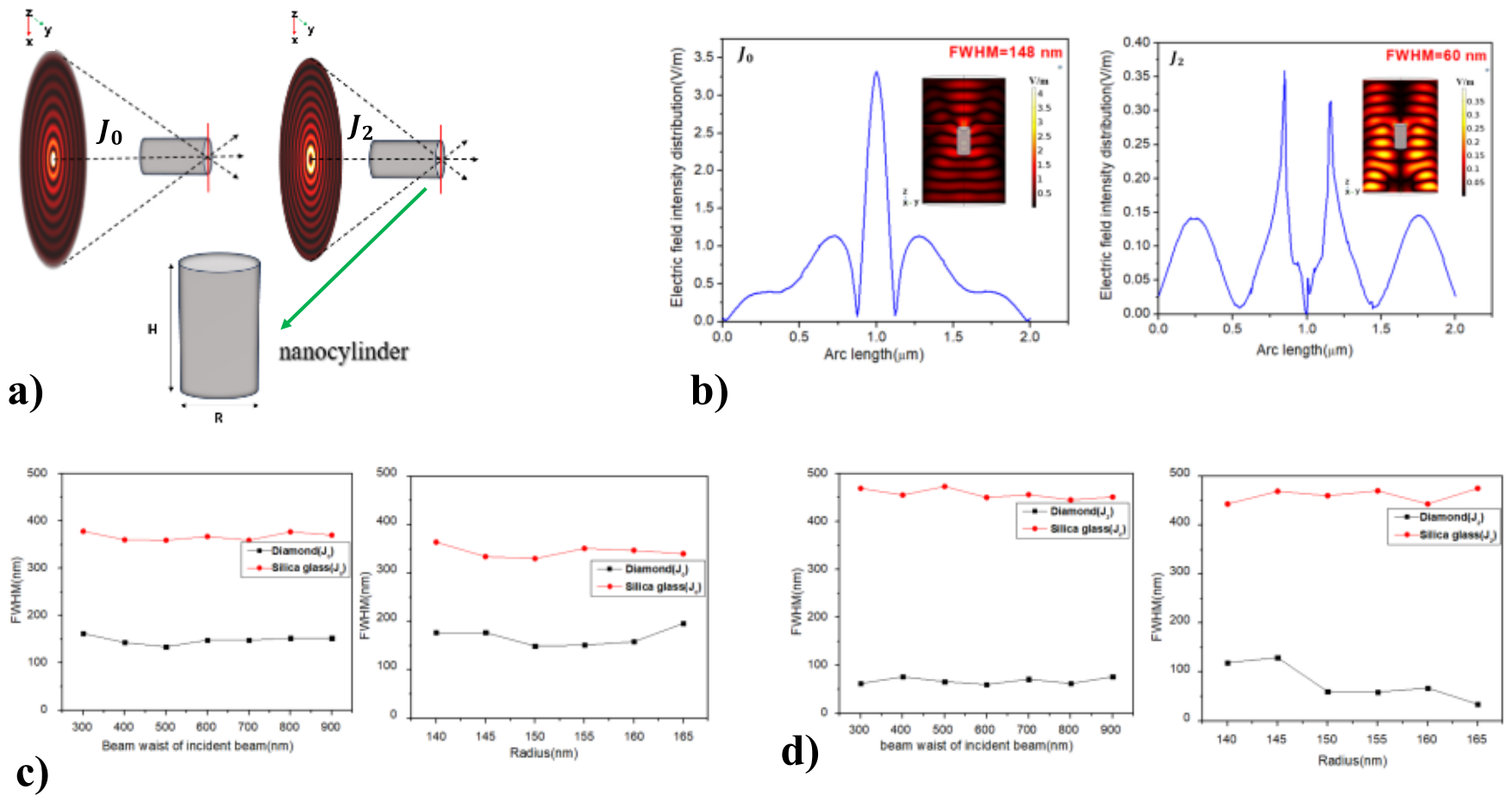

**Fig. 4** (a) Schematic representation of the zeroth and second order ($J_0$ and $J_2$) Bessel beam and geometrical description of nanocylinder. (b) Electric field distribution of the nanostructure and FWHM variation due to $J_0$ and $J_2$ are shown here (c) and (d) Dependence of the FWHM by different cylinder radius and beam waist are shown here to confirm the stability of the structure.

To systematically evaluate the near-field spatial confinement and geometry-dependent energy flow of the system, numerical simulations of cylinder geometry are compared without the inclusion of nanocylinder inside the simulation region. In the absence of the nano-cylinder (Figure 5 (a)), the unperturbed incident beam forms a wide, diffraction-limited profile across the POI with a characteristic FWHM of 455 nm. Upon integrating the high-index nanocylinder, the sudden discontinuity in the local permittivity alters the electromagnetic boundary conditions. Rather than allowing the wavefront to propagate unhindered, the subwavelength dielectric interfaces actively intercept and redistribute the incoming field energy, compressing it into a deeply sub-diffraction spatial volume with an ultra-narrow FWHM of 60 nm. This physical confinement mechanism is explicitly confirmed by the calculated Poynting vector flow patterns (Figure 5 (b)), where the transverse components of the vector field reveal a distinct lateral convergence of power toward the optical axis. This inward energy flow drives strong localized scattering and field enhancement, concentrating the optical power directly at the sharp structural boundaries to feed and sustain the highly localized near-field hotspots.

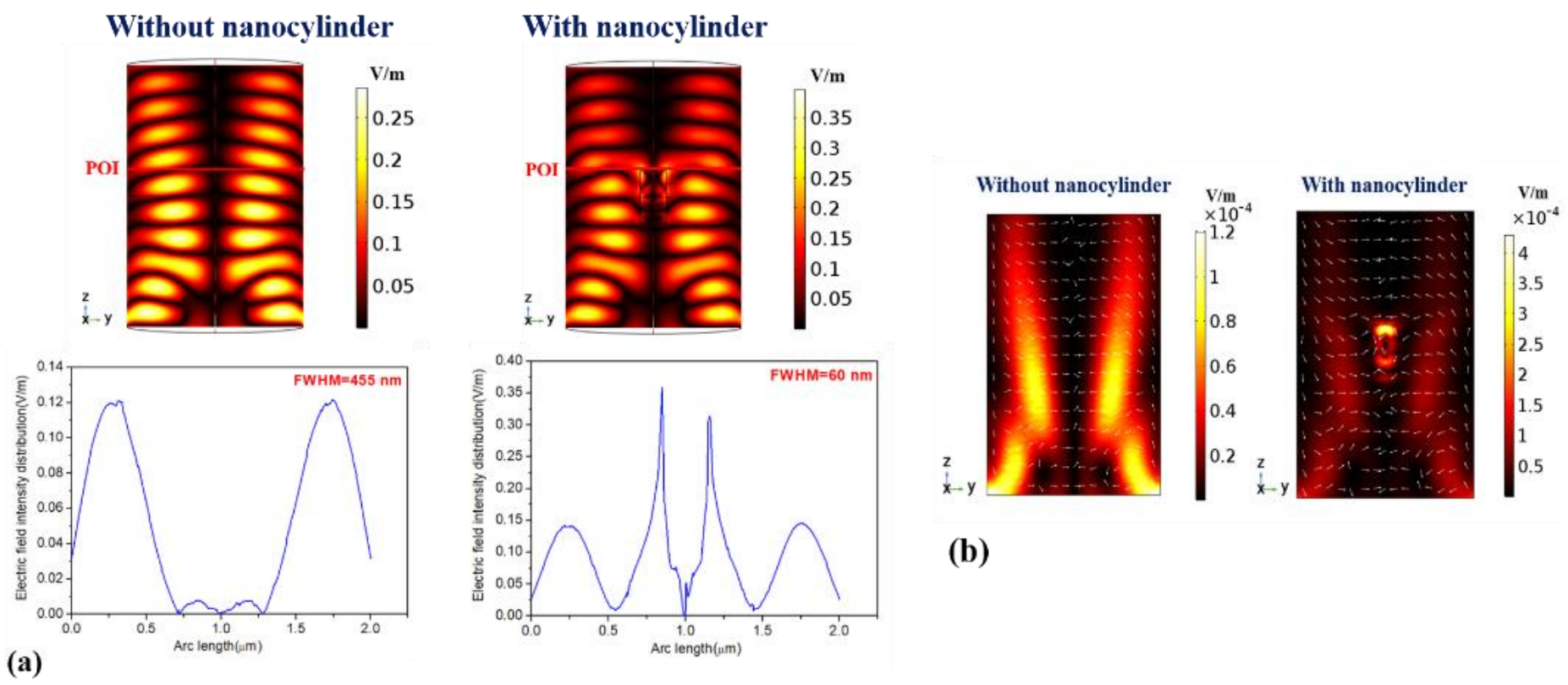

**Fig. 5.** Numerical analysis of near-field distribution and energy flow with and without the nanocylinder structure. (a) Electric field intensity distributions (V/m) in the yz plane along with the corresponding red

line indicates the POI, (b) Simulated Poynting vector flow patterns (white arrow surfaces) overlaid on the energy density profiles, illustrating the lateral convergence of optical power and the energy redistribution mechanism responsible for generating the highly localized hotspots at the structural boundaries.

## 3.4. Geometry-dependent confinement of Bessel beams in nanocone and nanopyramidal structures

Figure 6 represents the analysis of different geometrical/optical parameters of cone and pyramidal nanostructures under higher-order Bessel beam illumination. The near electric field distribution is strongly concentrated toward the tip of the nano-cone (figure 6(a)), driven by enhanced optical confinement, resulting in distinct near-field localization. For zeroth-order ($J_0$) illumination, optimized nanocone with R = 0.61 μm, H = 0.405 μm, yields a near field focal spot with FWHM =110 nm (figure 6(b)). A significant stable performance is observed for varying the cone angles (i.e. ~ 33°–35°) while the same characteristics is attained even with a wider beam waist variation (i.e. 500–900 nm) (Figure 6(c)). For both the cases, we have compared with a silica lens material. Owing to the higher refractive index contrast, diamond nanostructures exhibit significantly stronger confinement than silica. Under higher-order ($J_1$) Bessel illumination, the presence of a phase singularity produces a donut-shaped intensity profile, suppressing the central lobe while enhancing near field localization at the tip. Optimized geometrical parameters (R = 0.51 μm, H = 0.415 μm) result in improved focusing performance with FWHM value of 90 nm and a stability in the resolution is observed over different cone angles and beam waist variations which are shown in Figure 6(d). Overall, high-index nanostructures enable robust near-field enhancement and sub-diffraction-limited focusing.

We have also introduced a nano-pyramidal structure (Figure 6(e)), in which the length and width along the three-dimensional axes are systematically tuned to achieve near-field confinement. The pyramidal nanostructure is designed with a flat-tipped pyramidal geometry controlled by a specific ratio between top and bottom base lengths along xy cross-section. Here we have defined '**r**' as the bottom and top base length ratio along the xy axis (where **r** = ($l_1/r_1$) and ($l_2/r_2$)). This highly tapered configuration promotes strong electric field concentration, resulting in improved nanoscale light confinement. By tuning the value of r, the spatial confinement of the optical field can be precisely controlled, directly impacting the optimized resolution. The optimized structural parameters are ($l_1$ = 0.4 μm, $l_2$ = 0.7 μm, H = 0.405 μm) maintained across all incident Bessel modes during beam excitation.

For zeroth-order ($J_0$) excitation, a broader focal spot is observed that results in FWHM value of 139 nm with pronounced sidelobes, whereas higher-order modes exhibit distinctly different behaviour, $J_1$ yields a highly confined central peak with negligible sidelobes (Figure 6(f)). $J_1$ and $J_3$ modes generate an excellent near field localization with a FWHM value of 10 nm, and 20 nm respectively with suppressed sidelobes, attributed to phase-structured field redistribution with the optimized value of r=0.01 for both the cases and r=0.3 for zeroth-order Bessel beam. In contrast to cone geometries, the pyramidal structure introduces anisotropic field confinement and sharper edge-induced localization, enabling stronger subwavelength focusing performance. This structural advantage facilitates ultra-narrow focal spots that are not readily achievable in smoothly varying cone structures.

From fabrication perspective, a stable operation is observed for the $J_0$ mode which is shown in figure 6(g), where the FWHM exhibits only a slight variation with increasing ratio (r). We have varied the optimized parameter r from 0.15 to 0.35, indicating robust focusing performance and a broad fabrication tolerance**.** In contrast, the higher-order modes ($J_1$ and $J_3$) achieve significantly lower FWHM values at optimized smaller ratios (0-0.15) with the resolution increasing gradually as the ratio increases. This trend suggests that enhanced subwavelength confinement is obtained within a narrower geometrical regime, reflecting greater sensitivity to structural variations but providing superior focusing performance. Stable FWHM for wider incident beam waist variation (from 300 nm to 900 nm) is attained (Figure 6(h)). Overall, pyramidal nanostructure provides an enhanced control over near-field localization, enabling a better nanoscale imaging platform among subwavelength photonic structures.

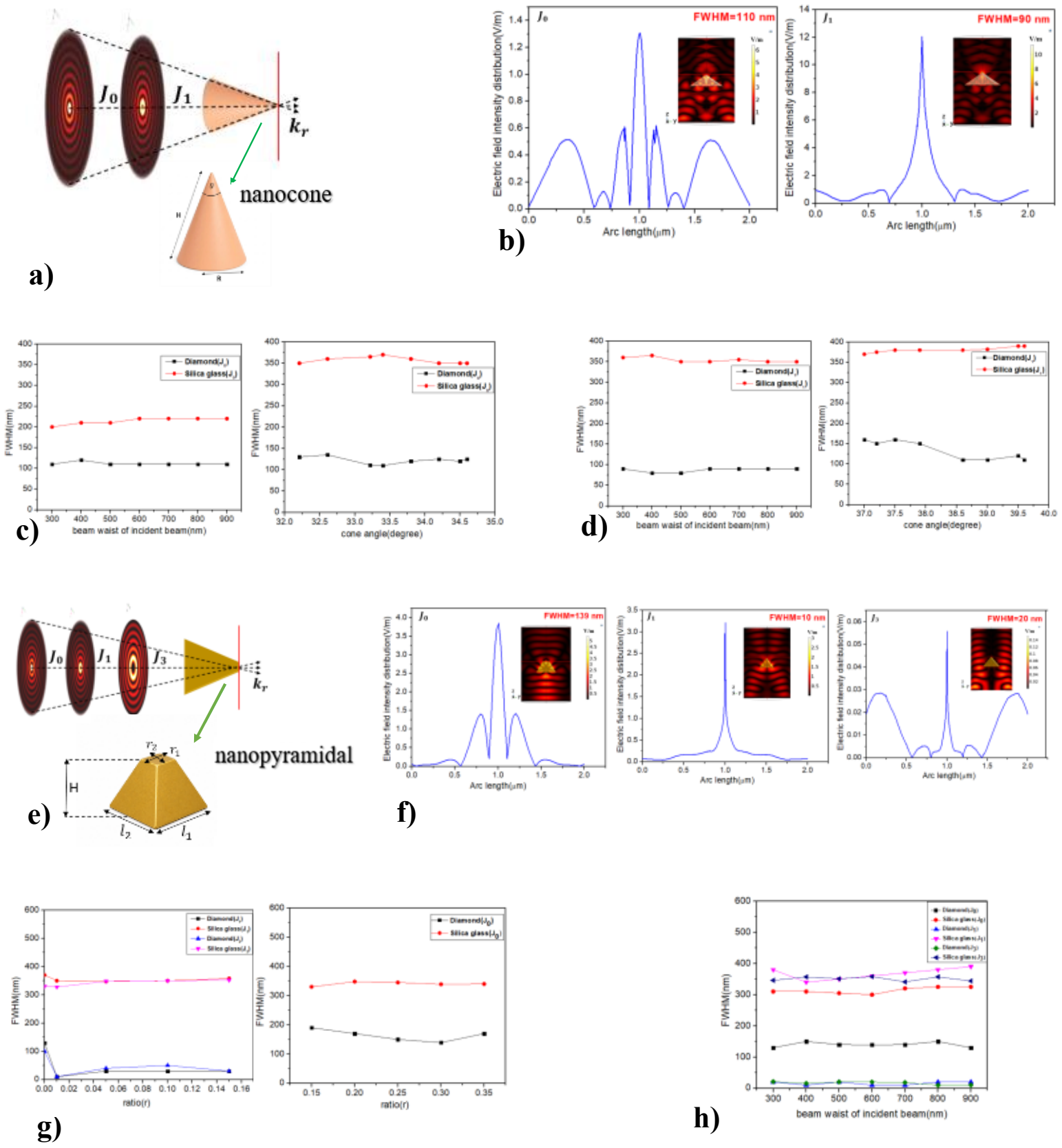

**Fig.6** (a), (e) Schematics of the beam propagation and geometry of cone and pyramidal nanostructures. (b), (f) Electric field intensity distributions for different Bessel beam orders with corresponding resolution analysis at plane of interest are shown. (c), (d) Effect of cone angle and beam waist on resolution in the nanocone structure are presented here, while (g), (h) show the influence of beam waist ($w_0$) and geometrical ratio (r) on the stability of the nanopyramidal structure for $J_0$, $J_1$, and $J_3$ orders.

To rigorously investigate the structural anisotropy of the system and its direct impact on spatial confinement, the localized near-field profiles are characterized along both orthogonal and transverse axes. The nanopyramidal geometry is intrinsically anisotropic, the induced local boundary conditions break azimuthal symmetry, causing the spatial widths of the localized optical fields to vary along different orientations. To accurately chart this behavior, both the x-directed FWHM_x and y-directed FWHM_y values are extracted from orthogonal one-dimensional line profiles across the apex interface, supported by high-resolution two-dimensional field intensity maps (Figure 7).

Under excitation by the $J_1$ mode (Figure 7 (a)), the system displays exceptional, tight near-field confinement that remains highly stable against geometric asymmetry. The 1D line profiles reveal a sharply localized confinement with an ultra-narrow FWHM_x = 9 nm along the (x)-axis and a marginally different FWHM_y = 10 nm along the y-axis. This near-uniform sub-10 nm localization demonstrates that the dominant near-field boundary effects at the pyramid apex can successfully sustain a nearly symmetric, deeply subwavelength focal spot despite the underlying geometric anisotropy. Conversely, when the system is driven by the higher-order $J_3$ mode (Figure 7 (b)), the structural anisotropy becomes far more pronounced in the field distribution. The complex spatial phase profile of the $J_3$ mode couples unevenly with the nanopyramidal facets, leading to a visible decoupling between the two axes. This results in relaxed field localization boundaries and noticeably disparate spatial spot sizes of FWHM_x = 26 nm and FWHM_y = 20 nm. This detailed dual-axis analysis directly demonstrates that tracking both FWHM_x and FWHM_y is essential, as it clearly reveals how the interplay between the incident mode order and the anisotropic nanopyramidal boundaries dictates the final two-dimensional topology and electric field confinement limits of the subwavelength hotspot.

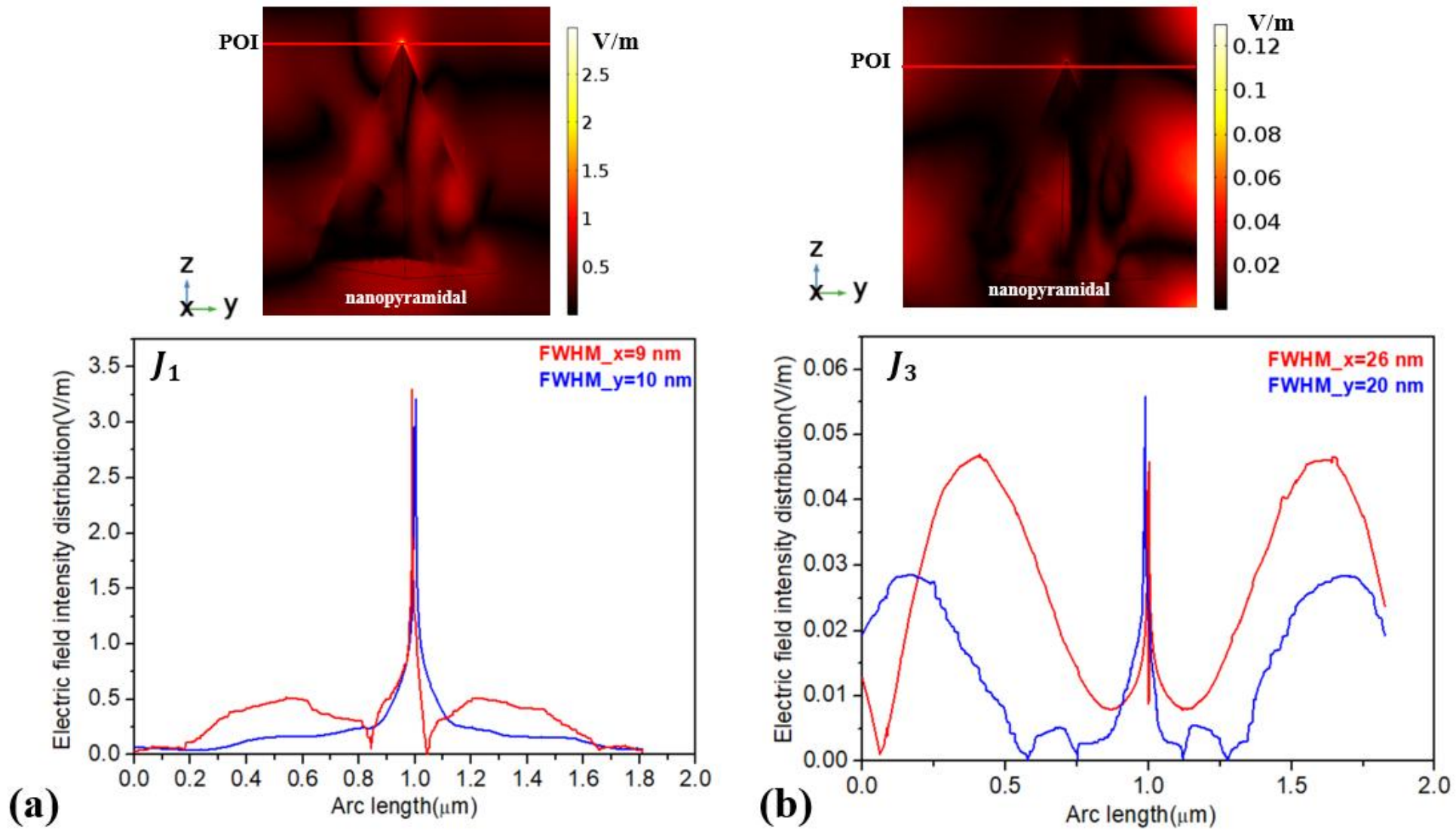

**Fig. 7 (a), (b)** Anisotropic near-field confinement and intensity profiles under $J_1$ and $J_3$ mode excitation. Top panels show 2D near-field maps at the nanopyramidal apex. Bottom panels show orthogonal 1D intensity profiles along the xz-axis (red) and yz-axis (blue), demonstrating geometry-dependent spatial subwavelength confinement limits.

## 3.5 Optical field confinement in solid immersion nanostructures driven by higher-order Bessel illumination

Figure 8(a) presents the performance of solid immersion lens (SIL)-based nanostructures under higher-order Bessel beam illumination. The propagation of a Bessel beam through a hemispherical SIL enables efficient coupling of its conical wavefront components with the curved dielectric interface, resulting in highly compressed focal spot along the longitudinal direction. Additionally, higher-order modes introduce controlled phase singularities that redistribute energy while maintaining strong localization, with $J_1$ providing optimal confinement. The geometrical and material parameters are defined as radius $R = 0.450\ \mu$m and refractive index $n = 2.405$ at an excitation wavelength of 532 nm. Consequently, spherical and ellipsoidal geometries fail to achieve efficient localization of optical energy beyond the diffraction limit, in contrast to hemispherical structures that enable stronger near-field confinement as shown in figure 8(b). This highlights the critical role of geometry-induced wavefront shaping in achieving subwavelength resolution. The analysis reveals that first-order Bessel beam ($J_1$) provides optimal near field confinement for the hemispherical nanostructures among the $J_0$, $J_1$, $J_2$ and $J_3$ modes, achieving a highly localized focal spot with minimal backscattering of field distribution. The corresponding FWHM is approximately 40 nm (figure 8(c)). In contrast, for lower refractive index materials such as silica glass (n=1.45) under identical conditions, the FWHM is significantly high as ~470 nm. Furthermore, variation in the structural configuration and beam order enables resolution beyond the classical diffraction limit. As shown in figure 8 (d), the zeroth-order Bessel beam ($J_0$) yields a FWHM of 66 nm, while $J_1$ achieves 40 nm. The second-order mode ($J_2$) produces a FWHM of 61 nm, and the third-order mode ($J_3$) results in 91 nm under an optimal radius at $R = 0.45\ \mu$m with different location spot. Notably, sidelobe intensities are significantly suppressed due to strong field localization.

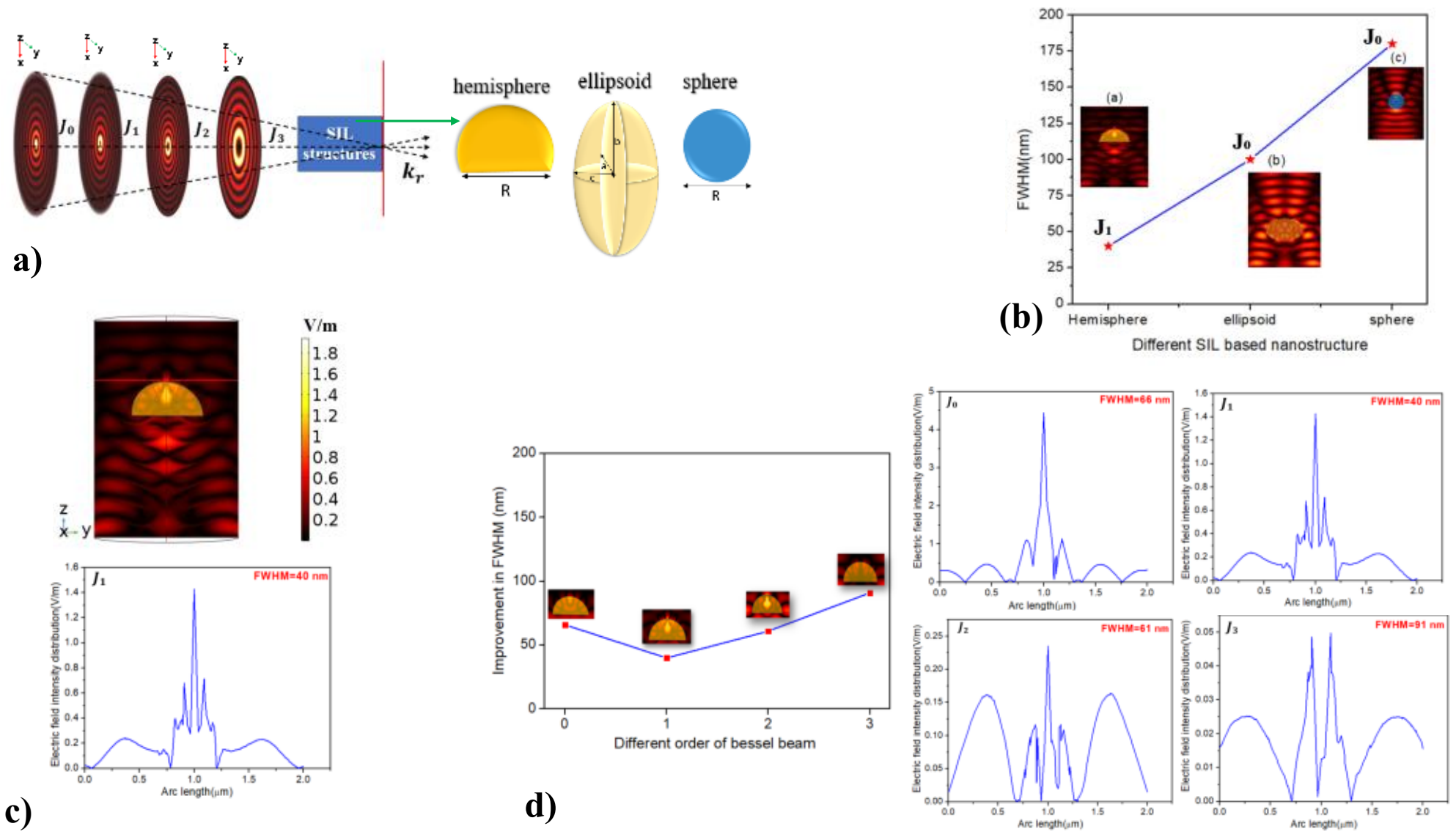

**Fig.8** (a) Methodology for generating near field focal spot under higher-order Bessel beam illumination in SIL based nanostructures, corresponding geometrical configurations are shown. (b) Comparative analysis of spatial resolution across different nanostructures (c), (d) Influence of higher-order Bessel modes $J_0$-$J_3$ on field confinement in a hemispherical nanostructure.

## 3.6 Comparative study of Bessel beam generation schemes in nanoscale propagation

Figure 9 depicts the resolution dependence at different plane of interest along the propagation axis under two different Bessel beam generation mechanisms. First, we present how the resolution changes with the distance after the beam passes through POI when we launch a Gaussian beam on diamond-based nano-axicon (Figure 9 (a)) and later we have calculated the same analysis when a Bessel beam is coupled with a nano-axicon and traverses along the propagation direction (Figure 9 (b)). The resolution trends shown in figures 9 (c) and 9 (d) directly reflect the longitudinal confinement, quantified by the FWHM of the intensity distribution extracted from the plane of interest to the extension of up to ~100 nm distance. The resolution is calculated over a propagation range extending from the tip of the nanostructure to ~100 nm along the optical axis, ensuring a consistent FWHM over a distance of ~100 nm for both excitation schemes. This can be extended to study to confirm in-depth resolution if we expand dimensions of the simulation parameters by an order of magnitude. In the Gaussian-to-Bessel conversion approach, the gradual increase in FWHM with propagation distance indicates a loss of high spatial frequency components, leading to reduced confinement and limited stability in the near-field localization. In contrast, direct Bessel beam illumination on nano-axicon maintains slightly better stable FWHM across the same propagation range, demonstrating superior stability and preservation of the non-diffracting field structure in the nanoscale regime. These observations confirm that the beam generation mechanism critically influences propagation stability and field confinement at subwavelength scales.

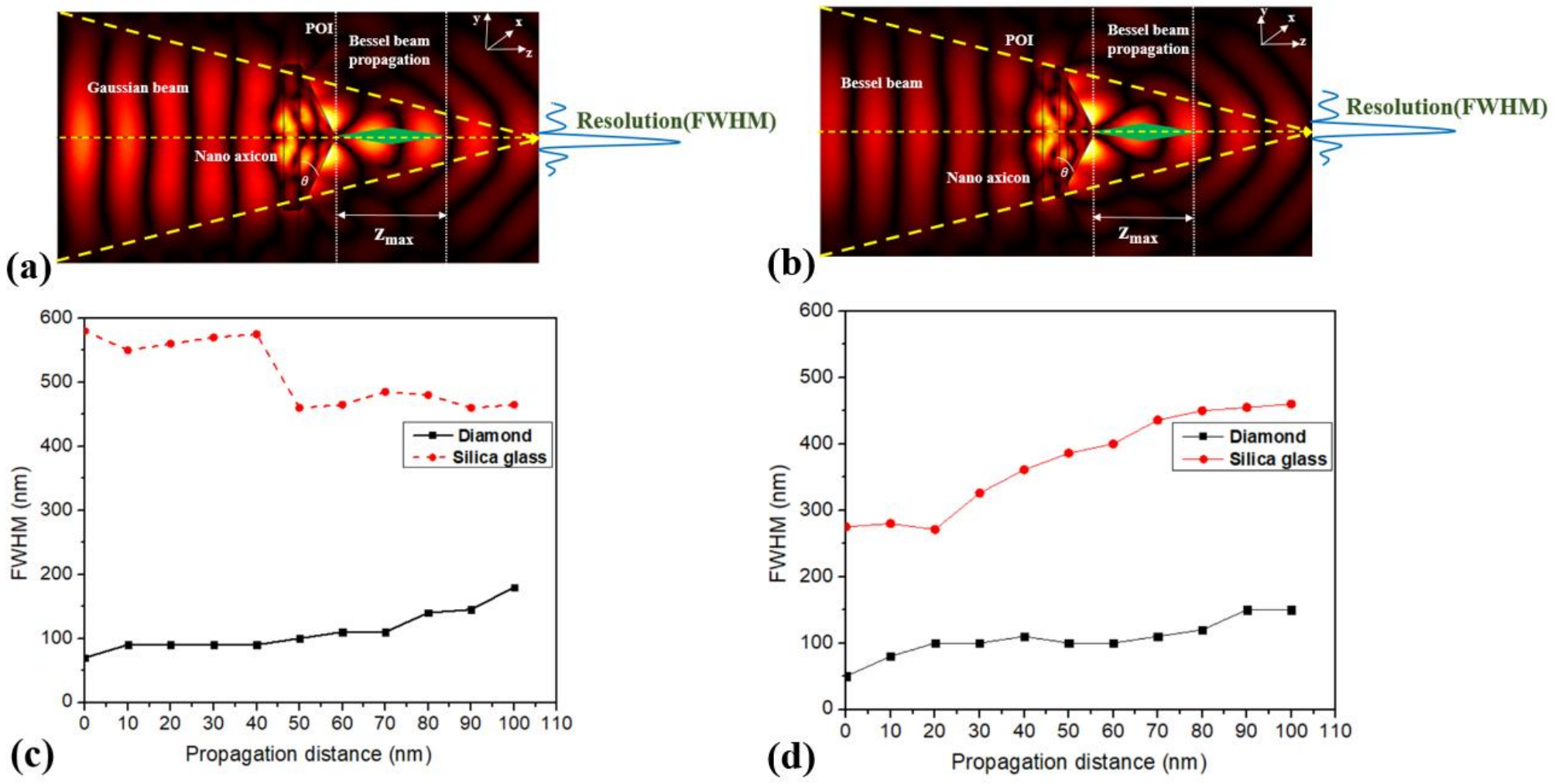

**Fig.9** Resolution versus propagation distance for two different Bessel beam generation schemes. (a) and (c) Gaussian-to-Bessel conversion via axicon lens shows increasing resolution due to loss of beam invariance. (b) and (d) Direct Bessel illumination maintains stable confinement. Diamond outperforms silica due to stronger field localization.

## 4. CONCLUSION

In this work, we demonstrate a unified framework for achieving subwavelength light confinement using optimized nanostructured geometries under higher-order Bessel beam illumination. Finite-element analysis reveals that these systems support robust near-field focusing with appreciable tolerance to structural and optical perturbations. Among the explored designs, the pyramidal geometry yields the strongest confinement, producing an ultra-narrow focal spot approaching ~λ/53 well beyond the conventional diffraction limit. Although residual sidelobes and fabrication limitations introduce geometry-dependent trade-offs, the overall simulation results confirm the practical viability of these techniques. More broadly, this work establishes a general route for integrating non-diffracting beam physics with nanophotonic designs, explore the opportunities for super-resolution 3D stack imaging, volumetric microscopy, optical manipulation and nanoscale laser lithography.

## REFERENCES

1. I. I. Smolyaninov, "Optical microscopy beyond the diffraction limit," HFSP J. **2**, 206–215 (2008).
2. D. Palounek, M. Vala, Ł. Bujak, I. Kopal, K. Jiříková, Y. Shaidiuk, and M. Piliarik, "Surpassing the diffraction limit in label-free optical microscopy," Nat. Photonics (2024). [Volume and page/article details to be verified.]
3. Z. Chen, A. Taflove, and V. Backman, "Photonic nanojet enhancement of backscattering of light by nanoparticles: a potential novel visible-light ultramicroscopy technique," Opt. Express **12**, 1214–1220 (2004).
4. Z. Liu, H. Lee, Y. Xiong, C. Sun, and X. Zhang, "Far-field optical hyperlens magnifying sub-diffraction-limited objects," Science **315,** 1686 (2007).
5. I. Golub, "Solid immersion axicon: maximizing nondiffracting or Bessel beam resolution," Opt. Lett. **31**, 1890–1892 (2006).
6. M. G. L. Gustafsson, "Surpassing the lateral resolution limit by a factor of two using structured illumination microscopy," J. Microsc. **198**, 82–87 (2000).
7. E. Betzig, G. H. Patterson, R. Sougrat, O. W. Lindwasser, S. Olenych, J. S. Bonifacino, M. W. Davidson, J. Lippincott-Schwartz, and H. F. Hess, "Imaging intracellular fluorescent proteins at nanometer resolution," Science **313**, 1642–1645 (2006).

8. M. J. Rust, M. Bates, and X. Zhuang, “Sub-diffraction-limit imaging by stochastic optical reconstruction microscopy (STORM),” Nat. Methods **3**, 793–795 (2006).
9. N. Fang, H. Lee, C. Sun, and X. Zhang, “Sub-diffraction-limited optical imaging with a silver superlens,” Science **308**, 534–537 (2005).
10. T. Umakoshi, “Near-field optical microscopy toward its applications for biological studies,” Biophys. Physicobiol. **20**(1), e200011 (2023).
11. S. W. Hell, “Far-field optical nanoscopy,” Science **316**, 1153–1158 (2007).
12. L. Novotny and B. Hecht, *Principles of Nano-Optics*, 2nd ed. (Cambridge University Press, 2012).
13. L. R. Hirsch, R. J. Stafford, J. A. Bankson, S. R. Sershen, B. Rivera, R. E. Price, J. D. Hazle, N. J. Halas, and J. L. West, “Nanoshell-mediated near-infrared thermal therapy of tumors under magnetic resonance guidance,” Proc. Natl. Acad. Sci. U.S.A. **100**, 13549–13554 (2003).
14. M. Ashrafian and L. Yousefi, "Subwavelength imaging using orbital angular momentum waves generated by metasurfaces," Sci. Rep. **15**, 20539 (2025).
15. L. Jiang, W. Zhang, H. Yuan, and X. Li, “Super resolution from pure/hybrid nanoscale solid immersion lenses under dark-field illumination,” Opt. Express **24**, 25224–25234 (2016).
16. S. Roy, J. Hu, and M. U. Momeen, “Superresolution technique beyond the diffraction limit under a structured beam via different optical nanostructures,” Sci. Rep. **16**, 66768 (2026). [Verify the article number and publication record.]
17. J. H. McLeod, “Axicon: a new type of optical element,” J. Opt. Soc. Am. **44**, 592–597 (1954).
18. D. Li, X. Wang, J. Ling, and Y. Yuan, “Planar efficient metasurface for generation of Bessel beam and super-resolution focusing,” Appl. Phys. A **127**, 232 (2021).
19. J. Durnin, J. J. Miceli, Jr., and J. H. Eberly, “Diffraction-free beams,” Phys. Rev. Lett. 58, 1499–1501 (1987).
20. L. Allen, M. W. Beijersbergen, R. J. C. Spreeuw, and J. P. Woerdman, “Orbital angular momentum of light and the transformation of Laguerre-Gaussian laser modes,” Phys. Rev. A **45**, 8185–8189 (1992).
21. A. E. Siegman, *Lasers* (University Science Books, Mill Valley, CA, 1986).

## Author Contributions Statement

M. U. M. conceptualized the idea. A. D. carry out the simulations and wrote the manuscript. All authors analyzed the results and edited the manuscript.

## Acknowledgments

The authors sincerely thank Dr. Subhankar Roy for the valuable scientific discussions during the execution of this work.

## Funding

This work was supported by Science and Engineering Research Board (now ANRF) under the grant no. SPG/2021/002905

## Disclosures

The authors declare no conflicts of interest.

## Data availability

Simulation concepts and protocols are explained in the manuscript. The related data are presented in the manuscript. If any further data are required, the authors are willing to provide.